\documentclass[a4paper]{amsart}
\usepackage{amssymb,latexsym}
\newtheorem{theorem}{Theorem}[section]

\usepackage{enumerate}
\begin{document}
\thispagestyle{empty}
\title{Secure Two-party Protocols for Point Inclusion Problem}
\maketitle
\begin{center}
Tony Thomas\\
Department of Mathematics\\
Korea Advanced Institute of Science and Technology\\
Daejeon, 305-701, Republic of Korea\\
Email: tonythomasiit@yahoo.com
\end{center}
\begin{abstract}
It is well known that, in theory, the general secure multi-party
computation problem is solvable using circuit evaluation protocols.
However, the communication complexity of the resulting protocols
depend on the size of the circuit that expresses the functionality
to be computed and hence can be impractical. Hence special
 solutions are needed for specific problems for efficiency reasons.
The point inclusion problem in computational geometry is a special
multiparty computation and has got many applications. Previous
protocols for the secure point inclusion problem are not adequate.
In this paper we modify some known solutions to the point inclusion
problem in computational geometry to the frame work of secure
two-party computation.
\end{abstract}
{\bf keywords}: multiparty computation, point inclusion problem,
computational geometry
\section{Introduction}
The rapid growth of networks has opened up tremendous opportunities
for cooperative computation, where the output depends on the private
inputs of several entities. These computations could even occur
between mutually untrusted entities or competitors. The problem is
trivial if the context allows to have a trusted entity that would
know the inputs from all the participants; however if the context
disallows this, then the techniques of secure multi-party
computation are used to provide useful solutions.\par Generally
speaking, a secure multi-party computation problem deals with
computing a function in a distributed network where each participant
holds one of the inputs, ensuring that no more information is
revealed to a participant in the computation than that can be
computed from that participant's input and output. The history of
the multi-party computation problem is extensive since it was
introduced by Yao ~\cite{yao1} and extended by Goldreich, Micali,
and Wigderson~\cite{gold2} and by many others. These works use a
similar methodology: each functionality $F$ is represented as a
Boolean circuit, and then the parties run a protocol for every gate
in the circuit. The protocols it generates depend on the size of the
circuit. This size depends on the size of the input and on the
complexity of expressing F as a circuit. If the functionality F is
complicated, using the circuit evaluation protocol will typically
not be practical. Therefore,  Goldreich~\cite{gold} pointed out that
 using the solutions from these general results for special cases of
 multi-party computation could be impractical; special
 efficient solutions should be developed for specific problems.
 This is the motivation for seeking  solutions to specific
cooperative computational problems, in which the solutions are more
efficient than the general theoretical solutions. To this end some
problems such as comparing two private
numbers~\cite{fisch,yong,ioa,lin}, privacy preserving data
mining~\cite{lind,agrawal}, comparing information~\cite{fagin},
privacy preserving geometric computation~\cite{du2}, privacy
preserving cooperative scientific computation~\cite{goldwasser,du1},
privacy preserving database query~\cite{du}, privacy preserving
auction~\cite{cachin}, privacy preserving statistical
analysis~\cite{du3,du4}, privacy preserving set
operations~\cite{lea} have been investigated.
\par In secure
multi-party computational geometry we seek secure protocols for
several geometric problem like point inclusion problem, intersection
of two shapes, range searching problem etc where the data is shared
by two or more entities. In this paper, we construct secure
two-party protocols for the point inclusion problem in star-shaped
domains and more complex polygonal domains. Here, one entity Alice
has a point $M$, and Bob has a polygon $P$. Their aim is to
determine whether $M$ is inside $P$, or not without revealing to
each other their private inputs.\par We outline the related work in
Section 2. In Section 3, we introduce our adversary models as well
as the cryptographic tools used in the subsequent sections. In
sections 4, we study the point inclusion problem in star-shaped
domains and in Section 5, we consider more general polygonal
domains. The paper concludes with some remarks in Section 6.
\section{Related Work}
The secure multiparty computational geometry has got wide
applications in the fields of military, computer graphics etc. The
study of secure multiparty computational geometry was initiated by
Atallah {\it et al.}~\cite{du2} with their work on secure point
inclusion problem and polygonal intersection problem. Their protocol
for the point inclusion problem is applicable to simple polygonal
domain and has complexity $O(n)$ where $n$ is the number of edges of
the polygon. Later Li {\it et al.}~\cite{li} studied the point
inclusion problem for circular domain. However, their solution is
not secure in the sense that each party gets additional information
regarding the location of the other party's object. Moreover, their
solution is highly inefficient. A more efficient protocol for the
point inclusion problem in a circular domain was recently proposed
by Luo {\it et al.}~\cite{lu}.\par In this paper we consider the
point inclusion problem in a star-shaped domain and a more general
polygonal domain (can have several disconnected nested components).
Two protocols for the star shaped domain with round complexities
$O(n)$ and $O(\log n)$ respectively, and a protocols for more
general polygonal domain  with round complexity $O(n)$, where $n$ is
the number of vertices are given.
\section{Preliminaries}
In this section we state our security assumptions and list the
building block for our protocols.
\subsection{Security Assumption}
We assume that all parties are semihonest. A semi-honest party is
the one who follows the protocol correctly with the exception that
it keeps a record of all its intermediate computations and might
derive the other parties inputs from the record.\par
 The existing protocols listed below serve as important building
blocks for our protocols.
\subsection{Homomorphic Encryption Schemes}
An encryption scheme is homomorphic if for some operations $\oplus$
and $\otimes$, $E_{k}(x) \otimes E_{k}(y) = E_{k}(x \oplus y)$,
where $x$ and $y$ are two elements from the message space and $k$ is
the key. Many such systems exist, and examples include the systems
by Benaloh~\cite{ben}, Naccache and Stern~\cite{nac}, Okamoto and
Uchiyama~\cite{okamoto}, Paillier~\cite{paillier}, to mention a few.
A useful property of homomorphic encryption schemes is that an
addition operation can be conduced based on the encrypted data
without decrypting them.
\subsection{Yao's Millionaire Protocol}
 The purpose of this protocol is to
compare two private numbers and to determine which one is larger
without revealing the numbers. This  was first proposed by
Yao~\cite{yao1} and is referred as Yao's Millionaire Problem
(because two millionaires wish to know who is richer, without
revealing any other information about their net wealth). The early
cryptographic solution by Yao~\cite{yao1} uses an untrusted third
party and has communication complexity that is exponential in the
number of bits of the numbers involved. Cachin proposed a
solution~\cite{cachin} based on an untrusted third party that can
misbehave on its own (for the purpose of illegally obtaining
information about Alice's or Bob's private vectors) but does not
collude with either participant. The communication complexity of
Cachin's scheme is $O(l)$, where $l$ is the number of bits of each
input number. Recently many efficient protocols which do not need a
third party have been suggested by various
authors~\cite{fisch,yong}.
\subsection{Scalar Product Protocol}
Let Alice has a vector $X = (x_1,\dots,x_n)$ and Bob has a vector $Y
= (y_1,\dots,y_n)$. The scalar product protocol is to securely
compute the scalar (dot) product of $X$ and $Y$, given by $X.Y =
\sum\limits_{k=1}^{n}x_iy_i$.\par In ~\cite{du2} Du and Atallah
considered a slightly different and more general form of the scalar
product protocol in which Alice has the vector $X$ and Bob has the
vector $Y$, and the goal of the protocol is for Alice (but not Bob)
to get $X.Y + V$ where $V$ is random and known to Bob only. Their
protocols can be easily modified to work for the version of the
problem where the random $V$ is given ahead of time as part of Bob's
data (the special case $V = 0$ puts us back to the usual scalar
product definition). They had developed two protocols for it. Secure
protocols for the scalar product problem can be found in
~\cite{du2,goe,yang}.
\section{Point Inclusion in Star-shaped Domain}In this
section, we study the point inclusion problem in a star-shaped polygonal domain.\\\\
{\bf Problem}: Let Alice has a point $M$ and Bob has a star-shaped
polygon $P$ with vertices  $P_i$,  for $1 \leq i \leq n$, where the
vertices are named in the anticlockwise direction. Alice and Bob
want to securely check
 whether $M$ lies inside (including boundary) $P$ or not.\\\par
Since $P$ is a star-shaped polygon, it contains a point $Q$ such the
line segments joining $Q$ to $P_i$ for $1 \leq i \leq n$ lies
entirely in $P$.
We have the following algorithm for point inclusion from ~\cite{pre}.\\\\
 {\bf The Point Inclusion Protocol Without Privacy}\\
\begin{enumerate}
\item Determine by binary search the wedge in which $M$ lies. $M$ lies in the
wedge bounded by the rays $\overrightarrow{QP_i}$ and
$\overrightarrow{QP_{i+1}}$ if and only if the angle formed by $M$,
$Q$ and $P_i$ is a left turn and the angle formed by $M$, $Q$ and
$P_{i+1}$ is a right turn.
\item Once $P_i$ and $P_{i+1}$ are found, then $M$ is internal if
and only if the angle formed by $P_i$, $P_{i+1}$ and $M$ is a left
turn.
\end{enumerate}
\vspace*{1cm}
\begin{theorem}~\cite{pre}
The inclusion question can be answered in $O(\log n)$ time, given
$O(n)$ space and $O(n)$ processing time.
\end{theorem}
To decide whether the angle $\angle P_1P_2P_3$ is a right or left
turn corresponds to evaluating a $3 \times 3$ determinant in the
points' coordinates. Let $P_i = (a_i,b_i)$ for $1 \leq i \leq 3$.
The determinant
\[ D(P_1,P_2,P_3) = \left|
  \begin{array}{ccc}
    a_1 & b_1 & 1 \\
    a_2 & b_2 & 1 \\
    a_3 & b_3 & 1 \\
  \end{array}
\right|
\]
gives twice the signed area of the triangle $\triangle P_1P_2P_3$,
where the sign is $+$ if and only if $(P_1,P_2,P_3)$ forms a
counterclockwise cycle. \par Let the coordinates of $M$ be $(a, b)$
and that of $Q$ be $(s, t)$ with respect to some coordinate system
known to both Alice and Bob. Now Bob chooses a new coordinate system
with origin
 at $Q$ and axes parallel to the original axes. Let the co-ordinates
 of $P_i$ with respect to the new coordinate axes be $(a_i, b_i)$ for
  $1 \leq i \leq n$. The new coordinates of $M$ becomes $(a -
  s, a -t)$. Now the angle $\angle MQP_i$ is a right turn or
  left turn according as the determinant
\[
D(M,Q,P_i) = \left|
  \begin{array}{ccc}
    a - s & b - t & 1 \\
    0 & 0 & 1 \\
    a_i & b_i & 1 \\
  \end{array}
\right|
\]
is positive or negative. For $1 \leq i \leq n$, let $A = (a, b, 1)$,
$B_i = (-b_i, a_i, sb_i-ta_i)$ and $C_i = ((b_i- b_{i+1}), -(a_i -
a_{i+1}), -s(b_i-b_{i+1})+t(a_i-a_{i+1})+(a_ib_{i+1}-b_ia_{i+1})$.
Now we have,
\begin{eqnarray*}
D(M,Q,P_i)& = &-(a -s)b_i + (b - t)a_i = - a b_i+ b a_i +
(sb_i-ta_i)\\& = &(a, b, 1).(-b_i, a_i,
sb_i-ta_i) = A.B_i.\\
 D(P_i,P_{i+1},M)& = &(a -s)(b_i-b_{i+1})
- (b -
t)(a_i-a_{i+1}) +(a_ib_{i+1}-b_ia_{i+1}) \\
&=& a (b_i- b_{i+1})- b (a_i - a_{i+1})
-s(b_i-b_{i+1})+t(a_i-a_{i+1})\\&&+ (a_ib_{i+1}-b_ia_{i+1})\\
& = &(a, b, 1). ((b_i- b_{i+1}), -(a_i - a_{i+1}),
-s(b_i-b_{i+1})+t(a_i-a_{i+1})\\&&+(a_ib_{i+1}-b_ia_{i+1}) = A.C_i.
\end{eqnarray*}
The point $M$ lies in the wedge bounded by the rays
$\overrightarrow{QP_i}$ and $\overrightarrow{QP_{i+1}}$ if and only
if $A.B_i \leq 0$ and $A. B_{i+1} \geq 0$ and if it happens to lie
in that wedge, it lies inside the polygon if and only if $A.C_i \leq
0$. Note that Alice has the vector $A$ and Bob has the vectors $B_i$
and $C_i$ for $1 \leq i \leq n$.
We now give the corresponding secure protocol for the point inclusion problem.\\\\
{\bf The Secure Point Inclusion Protocol 4.1}\\
\begin{enumerate}
\item For $i = 1,\dots, n$, Alice and Bob do the following:\\
\begin{enumerate}
\item Bob computes $B_i$, $C_i$ and chooses at random $V_i$ and $W_i$.
\item Alice engages in two secure scalar product protocols with Bob and gets $U_i = A.B_i +
V_i$ and $Z_i = A.C_i + W_i$.
\item Alice compares $U_i$
with $V_i$ and $Z_i$ with $W_i$ using millionaire protocol with
Bob.\\
\end{enumerate}
\item Alice identifies the index, $i = j$ at which $U_j < V_j$ and $U_{j+1} > V_{j+1}$.
\item Alice looks at the millionaire protocol output for the pair
$Z_j$ and $W_j$. If $Z_j$ was smaller than $W_j$ then the point is
inside else it is outside.
\item Alice communicates the result to Bob.\\
\end{enumerate}
{\bf Analysis of the Protocol 4.1}
\begin{theorem}
The Protocol 4.1 is correct, secure and has round complexity $O(n)$.
\end{theorem}
\begin{proof}
{\bf Correctness}: Using the millionaire protocol, in Step 2 Alice
 identifies the wedge in which the point
 $M$ lies and in Step 3 she checks whether the point $M$ lies inside
 or outside the polygon. The correctness of the protocol
follows from the correctness of the corresponding insecure protocol.   \\\\
{\bf Security}: The security of the protocol immediately follows
from the privacy of the secure scalar product protocol and that of
the secure protocol for the millionaire problem. Also, Alice does
not reveal to Bob the wedge in which $M$ lies, and so Bob  will not
get
any idea about the location of the point $M$. \\\\
 {\bf Round Complexity}:
It is easy to see that the round complexity of the protocol is
$O(n)$.
\end{proof}\vspace*{.5cm}
{\bf Using Binary Search to Reduce Round Complexity}\\\\
Now, we will incorporate binary search in the above protocol to
reduce its round complexity to $O(\log n)$.\par Let $E$ be a
homomorphic commutative encryption scheme. That is if $(E_A, D_A)$
and $(E_B, D_B)$ be the encryption and decryption pairs of Alice and
Bob corresponding to their keys and let $E = E_A$ or $E_B$,
then\\
\begin{enumerate}
\item $E_A(E_B(x)) = E_B(E_A(x))$;
\item $E(x)*E(y) = E(x.y)$.\\
\end{enumerate}
Given $U = (u_1,\dots, u_n)$, let $E(S) = (E(u_1),\dots, E(u_n))$.
We now give the modified
secure protocol for the point inclusion problem.\\\\
{\bf The Secure Point Inclusion Protocol 4.2}\\
\begin{enumerate}
\item For $1 \leq i \leq n$, Bob computes  $b_i = E_B(B _i)$, and $c_i = E_B(C_i)$.
 \item Bob sends $(b_{1},\dots,b_{n})$ and $(c_{1},\dots,c_{n})$ to Alice.
\item Alice picks an $r$ randomly such that $1 < r <n$ and cyclically rotates
the lists obtained from Bob by $r$ positions to get $(b_{1+r},\dots,
b_{r})$ and $(c_{1+r},\dots, c_{r})$.
\item Alice sends $(E_{A}(b_{1+r}),\dots,
E_{A}(b_{r}))$ and $(E_{A}(c_{1+r}),\dots, E_{A}(c_{r}))$ to Bob.
\item Bob decrypts the list obtained from Alice with his private key $D_B$
and obtains\begin{eqnarray*}
 (D_B(E_{A}(b_{1+r})),\dots,
D_B(E_{A}(b_{r}))) &=& (E_A(B_{1+r}),\dots,E_A(B_r)),\\
(D_B(E_{A}(c_{1+r})),\dots, D_B(E_{A}(c_{r}))) &=&
(E_A(C_{1+r}),\dots,E_A(C_r)).
\end{eqnarray*}
\item Alice computes  $E_A(A)$.
\item Alice identifies the index, $i = j$ for which $A.B_j < 0$ and $A.B_{j+1} > 0$
using the following sub protocol in the binary search.\\
\begin{enumerate}
\item For each index $k$ Alice picks up in the binary search, Bob picks a
random $r_k > 0$ encrypts with his key and sends Alice $E_B(r_k)$.
\item Alice encrypts with her key and sends back to Bob $E_A(E_B(r_k))$.
\item Bob decrypts and obtains $D_B(E_A(E_B(r_k))) = E_A(r_k)$.
\item Bob computes $E_B(r_k)* E_B(B_k) = E_B(r_kB_k)$.
\item Alice engages in a secure scalar product protocol with Bob and
obtains\\ $E_A(A)* E_A(r_kB_k) = E_A(r_k(A.B_k))$.
\item Alice decrypts and obtains $D_A(E_A(r_k(A.B_k)) = r_k(A.B_k)$
and checks whether it is positive or not.\\
\end{enumerate}
\item Alice checks whether $A.D_j$ is negative or positive using a similar sub protocol as in
 Step 7. If it is negative, the point is inside else it
is outside.
\item Alice communicates the result to Bob.\\
\end{enumerate}
{\bf Analysis of the Protocol 4.2}
\begin{theorem}
The Protocol 4.2 is correct, secure and has round complexity $O(\log
n)$.
\end{theorem}
\begin{proof}
{\bf Correctness}: It is clear that, in Step 5, Bob gets the
encryption of the vectors $B_i$ and $C_i$ for $1 \leq i \leq n$ with
the key of Alice. For each index $k$ occurring in the binary search,
Alice has $E_A(A)$ and Bob has $E_A(r_kB_k)$. Using the secure
scalar product protocol she obtains $E_A(A) * E_A(r_kB_k)$, which is
equal to $E_A(r_kA.B_k)$ from the homomorphic property of the
encryption scheme. By decryption using her private key Alice gets
$r_k(A.B_k)$ and she can check whether $A.B_k \geq 0$, since $r_k >
0$. Thus Alice can identify the wedge in which the point $M$ lies.
Similarly, once the wedge is identified, she can check whether the
point lies inside the polygon or not. Thus the correctness of the
protocol follows
from the correctness of the corresponding insecure protocol.  \\\\
{\bf Security}: Since Bob is sending $B_i$ and $C_i$ for $1 \leq i
\leq n$, after encryption with his key, Alice will not get any
information about the private data of Bob. Since Alice rotates the
list of $B_i$ and $C_i$ after masking with her key, Bob will not get
any idea of the specific $B_i$ and $C_i$ Alice is using in the
binary search in Step 7. Hence, Bob will not get any idea of the
wedge in which the point $M$ lies. The privacy of the secure scalar
product protocol guarantees the privacy of the individual inputs
during the scalar product computation in Step 7 and Step 8. Also
since $r_k$ is random known only to Bob, the only information Alice
can get from the
scalar product is its sign.\\\\
 {\bf Round Complexity}: As Alice is using binary search in the identification
 of the wedge in which the point $M$ lies, it is clear that the
round complexity of the protocol is $O(\log n)$, since the
complexity of the binary search is $O(\log n)$.
\end{proof}

\section{Point Inclusion in More General Polygonal Domain}In this
section, we consider an algorithm for the point inclusion problem
for a  more general polygonal domain given in ~\cite{fran}. This
domain is more general
than any of the domains so far considered in the context of secure point inclusion problem. \\\\
 {\bf Problem}: Alice has a point $M$ and Bob has a polygon $P$ that may have multiple disconnected nested
components, with vertices $P_1,\dots,P_n$. Alice and Bob wants to
securely check
 whether $M$ lies inside (including boundary) $P$ or not.\\\par
The {\it characteristic function}, $\chi(M)$ of the polygon $P$ is
defined as,
\[ \chi(M) = \begin{cases}1 ~~\text{if}~~M~~\text{lies on or inside}~~ P;\\
0~~\text{otherwise},
\end{cases}
\]
where $M \in \mathbb{R}^2$. Let $0 < \theta < 2\pi$, be the included
angle (edges swept inside the polygon) at a vertex $V$. The
extension to $\infty$ in both directions of the edges incident on
the vertex $V$ divide the plane into $4$ wedges. If $\theta < \pi$
(convex vertex), there are two wedges with angle $\theta$ and two
wedges with angle $\pi - \theta$. We call the wedges with angle
$\theta$ as inner and those with angle $\pi- \theta$ as outer. If
$\theta > \pi$ (concave vertex), there are two wedges with angle
$2\pi - \theta$ and two wedges with angle $\theta - \pi$. In this
case, we call the wedges with angle $2\pi - \theta$ as inner and
those with angle $\theta - \pi$ as outer.\par We assume for
convenience that the point $M$ does not lie on any of the four rays
emanating from any of the vertices of the polygon. The case in which
$M$ lies on a ray can be easily handled separately. Now, the {\it
cross function}, $\rho_{_V}(M)$ of a point $M$ with respect to a
vertex $V$ of the polygon is defined as\\
\[ \rho_{_V}(M) = \begin{cases}\frac{1}{2}-\frac{\theta}{2\pi} ~~\text{if}
~~\theta < \pi~~\text{and}~~ M~~\text{is in an inner wedge};\\
-\frac{\theta}{2\pi} ~~\hspace*{.4cm}\text{if}
~~\theta < \pi~~\text{and}~~ M~~\text{is in an outer wedge};\\
\frac{\theta}{2\pi} -\frac{1}{2} ~~\text{if}
~~\theta > \pi~~\text{and}~~M~~\text{is in an inner wedge};\\
\frac{\theta}{2\pi} ~~\hspace*{.6cm}\text{if} ~~\theta >
\pi~~\text{and}~~ M~~\text{is in an outer wedge}.
\end{cases}\\\vspace*{1cm}
\]
\begin{theorem}~\cite{fran}
The characteristic function of the whole polygon is the sum of the
cross functions of its vertices. That is
\[
\chi(M) = \sum_{V = P_1}^{P_n}\rho_{_V}(M), ~~~~~\forall M \in
\mathbb{R}^{2}.
\]
\end{theorem}
Before we give the secure protocol for the point inclusion problem,
 we outline a way for Alice to securely identify whether her point
 lies in an inner or outer wedge corresponding to a vertex $V$.
  Bob chooses four points $V_1, V_2, V_3$ and $V_4$ on the four rays emanating from the vertex
  $V$. Without loss of generality let us suppose that
  $\overrightarrow{VV_1}$ and $\overrightarrow{VV_2}$ bound one
  inner wedge and  $\overrightarrow{VV_3}$ and
  $\overrightarrow{VV_4}$ bound the other one. Now Alice and Bob
  engages in a secure protocol (as described in the previous section) and Alice checks
  whether $M$ is inside any of these two wedges. If that is the
  cases $M$ is inside an inner wedge else $M$ is inside an outer
  wedge.\par
For $1 \leq i \leq n$, let $\theta_i$ be the included angle at the
vertex $P_i$. We now give a secure protocol for the point inclusion problem.\\\\
{\bf The Secure Point Inclusion Protocol 5.1}\\
\begin{enumerate}
\item Bob computes $\theta =
\sum\limits_{i=1}^{n} (-1)^{m_i}\frac{\theta_i}{2\pi}$, where $m_i =
0$ if $V_i$ is a convex vertex ($\theta_i < \pi$) and $m_i =1$,
otherwise.
\item For $1 \leq i \leq n$ Alice and Bob do the following.\\
\begin{enumerate}
\item For the vertex $V_i$, Alice checks whether $M$
lies inside an inner or outer wedge using the protocol described
above.
\item If the wedge is inner, Alice assigns $u_i=\frac{1}{2}$, else
she assigns $u_i=0$.
\item If the edge is convex Bob assigns $v_i =1$, else he assigns
$v_i= -1$.\\
\end{enumerate}
\item Alice assigns $U = (u_1,\dots,u_n)$.
\item Bob assigns  $V = (v_1,\dots,v_n)$.
\item Bob engages in a secure scalar product protocol with Alice and gets $U.V$.
\item Bob computes $\chi(M) = U.V + \theta$.
\item Bob communicates the result to Alice.
\end{enumerate}
\subsection{Analysis of the Protocol 5.1}
\begin{theorem}
The Protocol 5.1 is correct, secure and has round complexity $O(n)$.
\end{theorem}
\begin{proof}
{\bf Correctness}: Let $E_1$ be the set of convex vertices where the
point $M$ lies in an inner wedge, $E_2$ be the set of convex
vertices where the point $M$ lies in an outer wedge, $E_3$ be the
set of concave vertices where the point $M$ lies in an inner wedge
and $E_4$ be the set of concave vertices where the point $M$ lies in
an outer wedge. Then  we have, \begin{eqnarray*} \chi(M) &=& \sum_{V
= P_1}^{P_n}\rho_{_V}(M)\\& =& \sum_{V_i \in E_1 }\rho_{_{V_i}}(M) +
\sum_{V_i \in E_2 }\rho_{_{V_i}}(M) + \sum_{V_i \in E_3
}\rho_{_{V_i}}(M)
+ \sum_{V_i \in E_4 }\rho_{_{V_i}}(M)\\
& =& \sum_{V_i \in E_1 }\frac{1}{2}-\frac{\theta_i}{2\pi} +
\sum_{V_i \in E_2 }-\frac{\theta_i}{2\pi} + \sum_{V_i \in E_1
}\frac{\theta_i}{2\pi} -\frac{1}{2} + \sum_{V_i \in E_1
}\frac{\theta_i}{2\pi}\\
&=& \sum_{i=1}^{n}(-1)^{m_i}\frac{\theta_i}{2\pi} + U.V
\end{eqnarray*}
Thus Bob can compute $\chi(M)$ and hence the protocol is
correct.\\\\
{\bf Security}: The security of the protocol immediately follows
from the privacy of the secure scalar product protocol and that of
the secure protocol for the millionaire problem.\\\\
{\bf Round Complexity}: It is clear that the round complexity of the
protocol is $O(n)$.
\end{proof}
\section{Conclusion}
In this paper, we studied the point inclusion problem for polygons
in $2$ dimension. The Protocol 4.2 for the star shaped domain has
far better round complexity than the existing protocols. The
Protocol 5.1 for the general polygonal domains is applicable for a
large class of polygonal domains than the existing protocols. We
hope to extend these ideas to more general domains and to higher
dimensions.

\end{document}